\documentclass{ws-procs975x65}

\begin{document}

\title{Testing General Relativity and gravitational physics using the LARES satellite}

\author{Ignazio Ciufolini$^1$, Antonio Paolozzi$^2$, Erricos Pavlis$^3$, John Ries$^4$, Vahe Gurzadyan$^5$, Rolf Koenig$^6$,
Richard Matzner$^7$,  Roger Penrose$^8$, Giampiero Sindoni$^2$}

\address{1.Dipartimento di Ingegneria dell'Innovazione, University of Salento, and INFN, Lecce, Italy,\\2.Scuola di Ingegneria Aerospaziale and DIAEE, Sapienza Universit\`a, di Roma. Rome. Italy,\\3.Planetary Geodynamics Laboratory, NASA Goddard Space Flight Center, Greenbelt, MD, USA,\\4.Center for Space Research, The University of Texas at Austin, TX, USA,\\5.Center for Cosmology and Astrophysics, Alikhanian National Laboratory, Yerevan, Armenia,\\6.Helmholtz Centre Potsdam GFZ German Research Centre for Geosciences,\\7.Center for Relativity, The University of Texas at Austin, TX, USA,\\8.Mathematical Institute, University of Oxford, UK}

\begin{abstract}    
The discovery of the accelerating expansion of the Universe, thought to be driven by a mysterious form of `dark energy' constituting most of the Universe, has further revived the interest in testing Einstein's theory of General Relativity. At the very foundation of Einstein's theory is the geodesic motion of a small, structureless test-particle. Depending on the physical context, a star, planet or satellite can behave very nearly like a test-particle, so geodesic motion is used to calculate the advance of the perihelion of a planet's orbit, the dynamics of a binary pulsar system and of an Earth orbiting satellite. Verifying geodesic motion is then a test of paramount importance to General Relativity and other theories of fundamental physics. On the basis of the first few months of observations of the recently launched satellite LARES, its orbit shows the best agreement of any satellite with the test-particle motion predicted by General Relativity. That is, after modelling its known non-gravitational perturbations, the LARES orbit shows the smallest deviations from geodesic motion of any artificial satellite: its residual mean acceleration away from geodesic motion is less than $4\times 10^{-12} m/s^2$. LARES-type satellites can thus be used for accurate measurements and for tests of gravitational and fundamental physics. Already with only a few months of observation, LARES provides smaller scatter in the determination of several low-degree geopotential coefficients (Earth gravitational deviations from sphericity) than available from observations of any other satellite or combination of satellites.
\end{abstract}

\section{Introduction}
General Relativity is a fundamental concept for understanding the universe that we observe\cite{MisThor, Hawking, CiufWheel}. It describes one of the four fundamental interactions of nature, the gravitational interaction governing the dynamics of large-scale systems and bodies such as planets, stars, galaxies and the Universe, as well as our daily attraction towards the center of Earth. Einstein's theory is a very well verified description of gravity\cite{CiufWheel, Whill, Turyshev}. But it has encountered unexpected developments in observational cosmology. The study of distant supernovae led to a discovery that distant galaxies accelerate away from us\cite{Riess, Perlmut99}. Since then, this dark energy\cite{Perlmut03, Caldwell}, regarded as a new exotic physical substance that is accelerating the expansion of the Universe, is at the center of attention of physics. And careful studies of that expansion also imply the presence of a large fraction of invisible (dark) but normally attracting matter\cite{BerHooSi}; dark energy and dark matter together constitute approximately 96\% of the mass-energy of the Universe. 

Rather than assuming the existence of these exotic material sources, the accelerated expansion of the Universe might be explained by changing the fundamental description of gravity. Some theories of gravitation alternative to General Relativity can explain the acceleration\cite{DvaGaPo, Carroll}; others have been ruled out\cite{Reyes}. Because most such theories have been much less intensively studied than General Relativity, continuing careful experimental tests of all aspects of gravitational interaction are needed.

General Relativity explains the gravitational interaction as the curvature of spacetime generated by mass-energy and mass-energy currents via the Einstein field equations\cite{MisThor, Hawking, CiufWheel}. For example, the gravitational attraction of Earth on its Moon and artificial satellites is explained by General Relativity via the spacetime curvature generated by the Earth's mass. The motion of any test-body within the gravitational field of another massive body, e.g., the motion of a `small' satellite around Earth, is simply determined by a geodesic of spacetime with curvature generated by the massive body. The Moon and the Earth artificial satellites approximately follow geodesics of spacetime with orbital perturbations from an ideal geodesic path due their finite size and the non-gravitational forces acting on them. A timelike geodesic path (world-line) in spacetime's Lorentzian geometry is one that locally maximizes proper time, in analogy with the length-minimizing property of Euclidean straight lines. A test-particle that follows geodesic motion is an electrically neutral body, with negligible gravitational binding energy compared to its rest mass, negligible angular momentum and is small enough that the inhomogeneities of the gravitational field within its volume have a negligible effect on its motion. Furthermore, non-gravitational perturbations must not influence its motion.

Thus, geodesic motion is at the foundation of General Relativity and of any other theory where the gravitational interaction is described by spacetime curvature dynamically generated by mass-energy. Therefore, the creation of the best possible approximation for the free motion of a test particle, a spacetime geodesic, is a profound goal for experiments dedicated to the study of the spacetime geometry in the vicinity of a body, yielding high precision tests of General Relativity and constraints on alternative gravitational theories. 

Important issues are now addressed regarding the approximation to a geodesic that is provided by the motion of an actually extended body. In General Relativity\cite{Hartle, Rindler}, the problem of an extended body is subtle, due not only to the non-linearity of the equations of motion, but also to the need to deal with the internal structure of the compact body, constructed of continuous media, where kinetic variables and thermodynamic potentials are involved. Further, there may be intrinsically non-local effects arising from the internal structure of the extended body, such as tidal influences. Moreover, there are problems concerning the approximations that need to be made in order to describe a given extended body as a test particle moving along a geodesic. These problems are related to the fact that many of the common Newtonian gravitational concepts such as the `center of mass', `total mass' or `size' of an extended material body do not have well defined counterparts in General Relativity\cite{Ehlers}.

The Ehlers-Geroch theorem\cite{EhGer} (generalizing the result in \cite{GerJang}) attributes a geodesic $\gamma$ to the trajectory of an extended body with a small enough own gravitational field, if for a Lorentzian metric the Einstein tensor satisfies the so-called dominant energy condition\cite{Hawking}, this tensor being non-zero in some neighborhood of $\gamma$ and vanishing at its boundaries\cite{GenRel}. This theorem, asserting that {\it`small massive bodies move on near-geodesics'}, thus achieves a rigorous bridge from General Relativity to space experiments with `small' satellites which suggests a high level of suppression of non-gravitational and self-gravitational effects from the satellite's own small gravitational field. This enables us to consider the satellite's motion to be nearly geodesic and hence provides a genuine testing ground for General Relativity's effects.

Given the extreme weakness of the gravitational interaction with respect to the other interactions of nature, the space environment is the ideal laboratory to test gravitational and fundamental physics. However, in order to test gravitational physics, a satellite must behave as nearly as possible as a test-particle and must be as little as possible affected by non-gravitational perturbations such as radiation pressure and atmospheric drag. In addition, its position must be determined with extreme accuracy.

The best realization of an orbiting test particle is LARES (LAser RElativity Satellite), Figure 1, which was successfully launched on February $13^{th}$ 2012 into a medium altitude, circular Earth orbit\cite{CPPRKM}. Its position can be measured with very high accuracy using the Satellite Laser Ranging (SLR) technique. Short-duration laser pulses (with a typical width of 10 ps for the state-of-the-art systems) are emitted from lasers on Earth and then reflected back by retro-reflectors on the artificial satellites. By measuring the total round-trip travel time, it is possible to determine the instantaneous distance to the satellite with an accuracy of a few millimeters. There are a number of spherical laser ranged artificial satellites, among which are Starlette, Stella, LAGEOS (LAser GEOdynamics Satellite) and LAGEOS-2, Ajisai, Etalon-1 and Etalon-2. The tracking data collected by the SLR network, illustrated in Figure 2, are analyzed, organized and distributed by the International Laser Ranging Service (ILRS)\cite{PeDeBo}.

\begin{figure}[htbp]
  \centering
  \includegraphics[width=120mm]{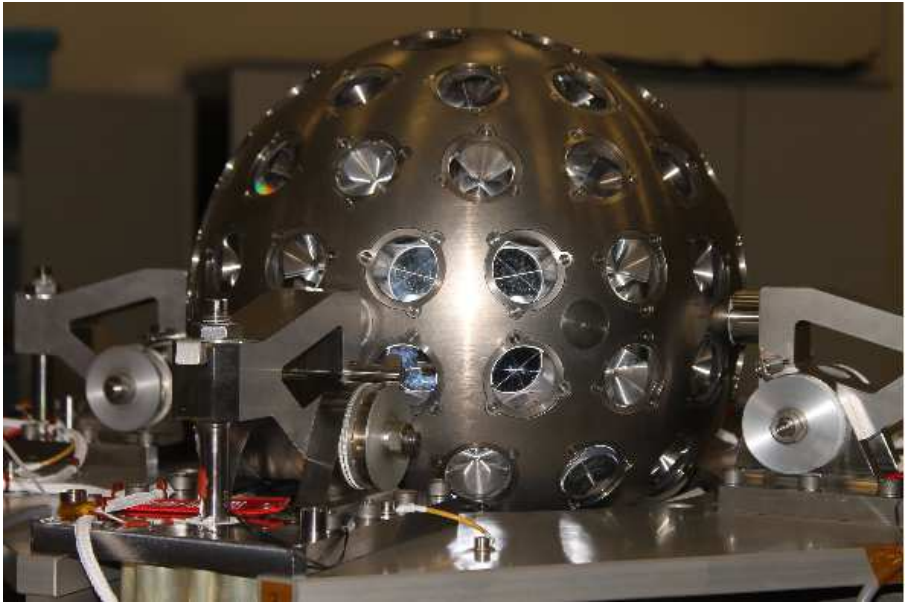}
  \label{weights}
  \caption{The LARES satellite on the separation system (courtesy of ASI).}
\end{figure}
 
 However, in order to test gravitational physics, we not only need to measure the position of a body with extreme accuracy, but we also need it to behave like a test-particle. In space, a test-particle can be realized in two ways: a small drag-free satellite or a small spacecraft with high density and an extremely small area to mass ratio. 

A drag-free satellite is an active spacecraft designed to reduce its non-gravitational orbital perturbations to extremely small values. It can be realized using a proof mass placed close to the satellite's center of mass, in a cavity that completely shields the proof mass. The satellite is then controlled by thrusters to follow the proof mass in such a way that the distance between the proof mass and the inner walls of the cavity does not change.Since the cavity is closed, the spherical proof mass is shielded from non-gravitational perturbations, such as radiation pressure and atmospheric drag. In the ideal case when the effects of other disturbing forces are negligible, the orbit of the proof mass, and therefore of the satellite itself, is determined only by gravitational forces. The only disturbing forces that can act on the proof mass arise from the satellite itself or from an interaction that can penetrate the shield. In the case of the Gravity Probe-B satellite, a mean residual acceleration of about $40\times 10^{-12} m/s^2$ was achieved\cite{Bencze}.

For a passive satellite (with no drag-free system), the key characteristic that determines the level of attenuation of the non-gravitational perturbations is the density, reflected by the ratio between its cross-sectional area and its mass. Until the launch of LARES, the two LAGEOS (LAser GEOdynamics Satellite) satellites had the smallest ratio of cross-sectional area to mass of any other artificial satellite\cite{Rindler} and were the best available test-particles. LAGEOS and LAGEOS-2 have an essentially identical structure but different orbits. They are passive, spherical satellites covered with retro-reflectors and made of heavy brass and aluminum. The mass of LAGEOS is 407 kg and that of LAGEOS-2 is 405.4 kg; their radius is 30 cm. The ratio of cross-sectional area to mass is approximately 0.0007$m^2/kg$ for both satellites. LAGEOS was launched in 1976 by NASA, and LAGEOS-2 by NASA and ASI (Italian Space Agency) in 1992. The semi-major axis of LAGEOS' orbit is approximately 12,270 km, the period is 3.76 h, the eccentricity is 0.004 and the inclination is 109.8$^o$. LAGEOS-2 is at a similar altitude but the orbit eccentricity is larger (0.014) and the inclination is lower (52.6$^o$). 

LARES (LAser RElativity Satellite), also from ASI, was launched with the qualification flight of VEGA, the new European Space Agency's (ESA's) launch vehicle developed by ELV (Avio-ASI). LARES is a spherical laser ranged satellite covered with 92 retro-reflectors with a radius of 18.2 cm. It is made of a tungsten alloy, with a total mass of 386.8 kg\cite{CPPRKM}, resulting in a cross-sectional area to mass ratio that is about 2.6 times smaller than LAGEOS. It has almost certainly the highest mean density of any object orbiting in the Solar System. It is currently very well observed by the ILRS stations that are located all over the world (Figure 2). The LARES orbital elements are: semi-major axis 7820 km, orbital eccentricity 0.0007 and orbital inclination 69.5$^o$. 

\begin{figure}[htbp]
  \centering
  \includegraphics[width=120mm]{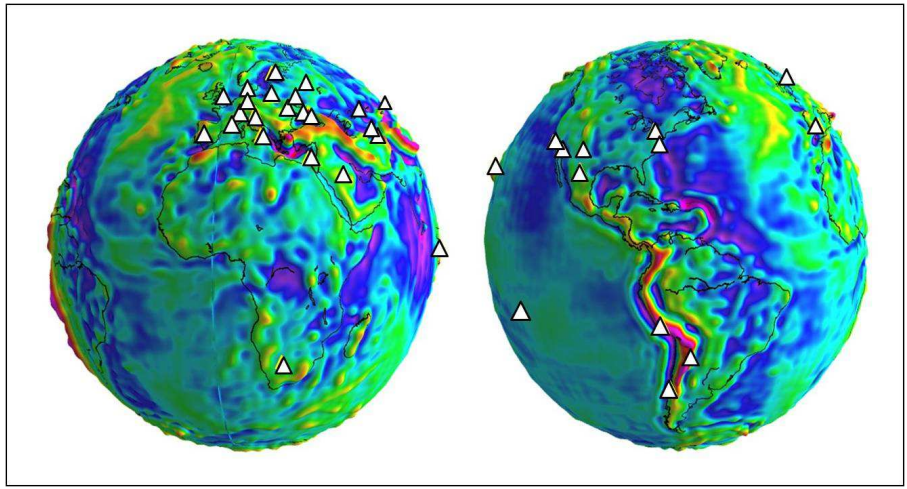}
  \label{weights}
  \caption{The current International Laser Ranging Service networkshown on a model of Earth's gravitational field obtained 
by the GRACE space mission\cite{LAGEOS85,RFKMN}}
\end{figure}

The extremely small cross-sectional area to mass ratio of LARES, which is smaller than that of any other artificial satellite, and its special structure, a solid-sphere with high thermal conductivity, ensure that the non-gravitational orbit perturbations are smaller than for any other satellite, in spite of its lower altitude compared to LAGEOS. This behavior has been experimentally confirmed using the first few months of laser ranging observations. 

We processed the LARES laser ranging data based on the first seven 15-day arcs using the orbital analysis and data reductionsystems UTOPIA of UT/CSR (Center for Space Research of The University of Texas at Austin), GEODYN II of NASA Goddard, and EPOS-OC of GFZ (Helmholtz Centre Potsdam GFZ German Research Centre for Geosciences). In one case using GEODYN, we estimated only a single residual along-track empirical acceleration, and in another case using UTOPIA, together with the residual acceleration, we estimated the degree-2 and -3 geopotential coefficients. The latter case was intended to take into account the lower altitude of LARES and thus the larger effect of gravity model errors on the orbit; the estimation of the gravity terms had little effect on the residual acceleration. For LARES, we modeled drag using its `best fit' drag coefficient $C_d$. The reflectivity coefficient, $C_\gamma$, was also the `best fit' value, which was found to be consistent with the measured surface properties and with similar spherical satellites. In all cases, state-of-the art satellite orbit dynamical models were employed, including all the general relativistic post-Newtonian corrections, GRACE-based mean gravity field models\cite{LAGEOS85, RFKMN}, modern models for the ocean and solid earth tides, as well as solar radiation pressure, Earth albedo and atmospheric drag\cite{Pavlis, Rubincam1}. No `thermal thrust'\cite{Rubincam2} models (described more fully below) were used. For the 105 days analyzed, both GEODYN and UTOPIA determined that the residual Pavlisalong-track accelerations for LARES were only about $0.4\times 10^{-12} m/s^2$, whereas for the two LAGEOS satellites, the acceleration residuals were $1-2\times 10^{-12} m/s^2$. As a further test, we fit the drag coefficient $C_d$ of LARES over the odd 15-days arcs and we then applied the mean $C_d$ obtained over these odd arcs to estimate the mean residual acceleration over the even arcs and vice-versa. We found again a residual acceleration of LARES of less than $0.4\times 10^{-12} m/s^2$. This is particularly impressive given that LARES is far lower in the Earth's atmosphere than LAGEOS.

The residual along-track accelerations of a satellite provide a measure of the level of suppression of its non-gravitational perturbations: atmospheric drag, solar and terrestrial radiation pressure and thermal-thrust effects. Atmospheric drag acts primarily along the satellite's velocity vector, while solar radiation pressure, terrestrial radiation pressure (the visible and infrared radiation from Earth) and thermal-thrust effects will all have some contribution along-track as well.The `classic' Yarkovsky effect on a spinning satellite is a thermal thrust resulting from the anisotropic temperature distribution over the satellite's surface caused by solar heating. In particular, a variation of this effect due to the Earth's infrared radiation is the Earth-Yarkovsky or Yarkovsky-Rubincam effect\cite{Rubincam2}. Infrared radiation from Earth is absorbed by the retro-reflectors; due to their thermal inertia and the rotation of the satellite, a latitudinal temperature gradient develops. The corresponding thermal radiation causes a significant along-track acceleration in the direction opposite to the satellite's motion. For LAGEOS and LAGEOS-2, this is estimated to be of the order of $1\times 10^{-12} m/s^2$. In addition to the Earth-Yarkovsky effect, there are also solar-Yarkovsky and asymmetric reflectivity effects that further perturb the LAGEOS orbits\cite{MVRE}; these also tend to be on the order of $1-2\times 10^{-12} m/s^2$. 

One of the reasons for the smaller residual acceleration of LARES with respect to LAGEOS, in spite of the lower altitude of LARES, is that the various thermal thrust effects should be much smaller on LARES than on LAGEOS. Indeed, LARES is much smaller than LAGEOS (18 cm radius versus 30 cm radius for LAGEOS), and it has higher thermal conductivity since it is a solid one-piece sphere. In contrast, LAGEOS is constructed from three separate pieces, thus decreasing the overall thermal conductivity properties. Furthermore, the effect of the thermal acceleration due to the retro-reflectors (which are the main source of the Earth and solar Yarkovsky effects) is also smaller because of the smaller cross-sectional-area to mass ratio of LARES and because the total surface area of the retro-reflectors is smaller on LARES (about 26\% of the total surface area) than on LAGEOS (about 43\%).

The effects of the residual unmodelled along-track acceleration on the orbits of LARES and LAGEOS are illustrated in Figure 3A where we plot the change in the distance from their `ideal' orbit caused by the unmodelled along-track accelerations. In Figure 3B, we show the effects of the residual unmodelled along-track acceleration on the orbits of LARES, LAGEOS and Starlette. The axis of ordinates may be thought of as representing an `ideal' reference world line of LARES, LAGEOS and Starlette, `ideal' in the sense that all of its orbital perturbations are known. Figures 3A and 3B show the unmodelled deviations from geodesic motion for LARES, LAGEOS and Starlette (once the known non-gravitational perturbations are removed, to the extent permitted by our current models) due to the unmodelled along-track accelerations. In these figures, we show the effect of a typical residual unmodelled along-track acceleration of $1\times 10^{-12} m/s^2$ for LAGEOS, $0.4\times 10^{-12} m/s^2$ for LARES and $40\times 10^{-12} m/s^2$ for Starlette. Since all the general relativistic post-Newtonian corrections were included in our orbital analyses, these figures show the level of agreement of the LARES and LAGEOS orbits with the geodesic motion predicted by General Relativity.

\begin{figure}[htbp]
  \includegraphics[width=130mm]{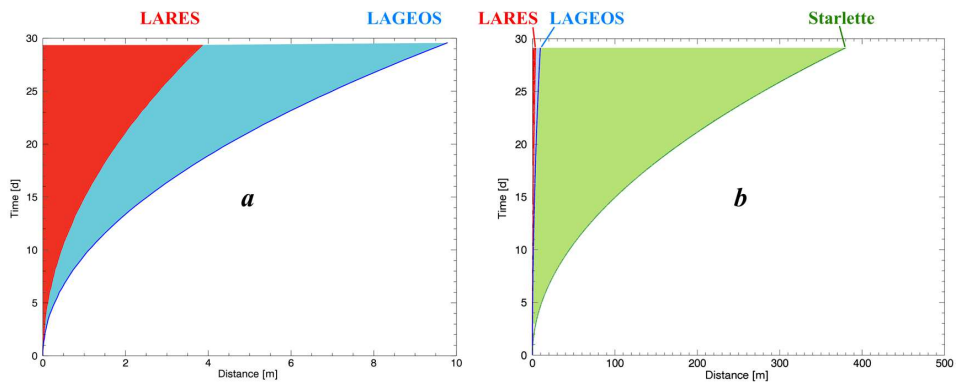}
  \label{weights}
  \caption{A. The red curve represents the change of distance between a ``test-particle'' following a spacetime geodesic, represented here by the axis of ordinates in a frame co-moving with the test-particle, and a similar particle perturbed by the average unmodelled along-track acceleration of the magnitude observed on the LARES satellite of approximately $0.4\times 10^{-12} m/s^2$. The blue curve represents the change of distance between a test-particle and a similar particle perturbed by an average along-track acceleration of the typical size of the unmodelled along-track acceleration observed on the LAGEOS satellites of the order of $1\times 10^{-12} m/s^2$. The axis of ordinates may be thought of to represent a spacetime geodesic followed by LARES or LAGEOS after removing all the known and unmodelled non-gravitational perturbations. Fig.3 B. Similar plots for LARES, LAGEOS and Starlette are shown. For Starlette, we used a typical residual acceleration of the order of $40\times 10^{-12} m/s^2$}
\end{figure}

It must be stressed that a residual unmodelled out-of-plane acceleration of the order of magnitude of the unmodelled along-track acceleration observed on LARES will produce an extremely small {\it secular} variation of the longitude of its node (the intersection of a satellite's orbital plane with the Earth's equatorial plane), i.e., of its orbital angular momentum. For example, by considering an out-of-plane acceleration with amplitude of $0.4\times 10^{-12} m/s^2$, its effect on the node of LARES would be many orders of magnitude smaller than the tiny secular drift of the node of LARES due to frame-dragging\cite{Ciufolini} of about 118 milliarcsec/yr. Therefore, LARES, together with the LAGEOS satellites, and with the determination of Earth's gravity field obtained by the GRACE mission, will be used to accurately measure the frame-dragging effect predicted by General Relativity, improving by perhaps an order of magnitude the accuracy of previous frame-dragging measurements by the LAGEOS satellites\cite{CiuPa, CPRKSP}. 

In fundamental physics, limits on certain possible low-energy consequences of string theory that may be related to dark energy and quintessence have been set using measurements of frame-dragging by the LAGEOS satellites\cite{SmErCaKa}. These limits will be improved with LARES. In addition to these fundamental physics tests, the LARES satellite will be an important geodetic contribution to the monitoring of the long-wavelength variations in Earth's gravity field due to mass redistribution due to the effects of global climate change. Until there are continuous GRACE-type gravity-monitoring missions, the SLR tracking to the `cannonball' satellites continue to provide a critical measure of the long-term long-wavelength gravity changes. Figure 4 illustrates the improved observability that LARES provides for the low-degree (long-wavelength) gravity terms (the degree 2 and 3 terms mentioned previously). The increased error with increasing degree reflects the signal attenuation that is a function of the satellite altitude, which is more severe for the higher LAGEOS satellites than for the lower LARES and Starlette satellites. While LARES is only slightly higher than Starlette (the Starlette altitude ranges from 800 to 1100 km compared to about 1440 km for LARES), LARES is able to resolve the low degree terms as well or better than Starlette, and much better than the LAGEOS satellites. In particular, for $C_{30}$, where the biasing effect of the thermal-thrust forces becomes important, LARES is considerably better than LAGEOS or LAGEOS-2. 

\begin{figure}[htbp]
  \centering
  \includegraphics[width=120mm]{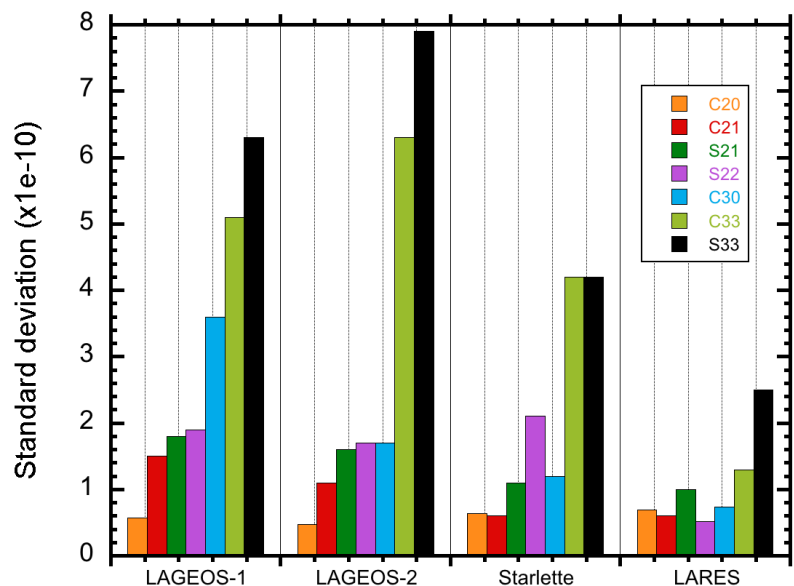}
  \label{weights}
  \caption{Statistics of the scatter in the determination of several low-degree geopotential coefficientsusing LAGEOS, LAGEOS-2, Starlette and LARES. }
\end{figure}

\section*{Acknowledgments}

In conclusion, LARES provides the best available test-particle in the Solar System for tests of gravitational physics and General Relativity, e.g., for the accurate measurement of frame-dragging and,after modelling its known non-gravitational perturbations,its orbit shows the best agreement of any satellitewith the geodesic motion predicted by General Relativity. LARES can also be used to set limits on other theories of fundamental physics and for a number of measurements in space geodesy and geodynamics. LARES-type satellites, placed in different orbits, could be used for a number of other tests of gravitational and fundamental physics. For example, tests of Brane-World theories, further limits on some low-energy consequences of string theory, possibly related to dark energy and quintessence, Yukawa-type deviations from the standard inverse square law for gravity and further measurements of the so-called Post-Newtonian parameters, testing General Relativity against alternative gravitational theories, could be carried out.


\begin{thebibliography}{99}
\bibitem{GenRel} In General Relativity the Lorentzian metric describes the curved spacetime geometry, where light speed is the limit for information transfer. The Einstein tensor locally describes mass-energy in the Einstein equations, and the dominant energy condition ensures its overall positivity.
\end{thebibliography}
\end{document}